\documentclass[12pt]{article}
\usepackage{graphicx,color}
\usepackage{hyperref}
\usepackage{amssymb,amsfonts,amsmath,cancel,cite,multirow}
\usepackage[capitalise]{cleveref}

\newcommand{\pdiff}[3]{
\if 1#1   \frac{\partial #2 }{\partial #3 }
\else  \frac{\partial^#1#2 }{\partial #3^#1 } \fi}

\begin{document}

\begin{titlepage}


\vskip 1.35cm
\begin{center}

{\large
\textbf{
Proton Decay in SUSY GUTs
}}
\vskip 1.2cm

Junji Hisano\\

\vskip 0.4cm

\textit{$^a$Kobayashi-Maskawa Institute for the Origin of Particles and the Universe (KMI),
Nagoya University, Nagoya 464-8602, Japan}

\vskip 1.5cm

\begin{abstract}
  I review proton decay in the SUSY SU(5) GUTs assuming the mini-split
  SUSY breaking model. In the mini-split SUSY breaking model, the
  squark and slepton masses are $O(10^{(2-3)})$~TeV while the gaugino
  masses are $O(1)$~TeV. As the result, even the minimal SUSY SU(5)
  GUT is still viable in the model. We present the motivation of the
  mini-split SUSY model and discuss the future prospects of proton
  decay searches in the SUSY SU(5) GUTs.
\end{abstract}

\end{center}
\end{titlepage}

\section{History of proton decay in GUTs}
Proton decay is a window for the Grand Unified Theories (GUTs)
\cite{GG,GQW,review}. GUTs have two aspects. The first one is the
unification of three forces in the Standard Model (SM), and another is
the unification of quarks and leptons. As the result, baryon ($B$) and
lepton ($L$) numbers are not conserved, and proton decay is predicted
in GUTs. Since proton decay is mediated by the GUT particles, it has
been argued that proton decay is a direct probe of the GUTs. The
minimal gauge group of GUTs is SU(5). When assuming quarks and
leptons, including the right-handed neutrinos, are unified into a GUT
multiplet for each generation, the gauge group is SO(10) or the
larger.

Proton decay has been being searched for since Georgi and Grashow
proposed the SU(5) GUTs and Georgi, Quinn and Weinberg evaluated
proton lifetime in the models in 1974 \cite{GG,GQW}. The $X$ boson is
introduced in the SU(5) GUT since the gauge group is extended from
SU(3)$_C\times$SU(2)$_L\times$U(1)$_Y$ in the SM. Integration of the
$X$ boson generates the baryon-number violating dimension-six
operators for proton decay. The main decay mode is $p\rightarrow \pi^0
{\rm e}^+$. The proton lifetime is proportional to the fourth power of
the $X$ boson mass ($M_X$). In 1980s, the large volume detectors were
constructed and operated to search for proton decay \cite{exp80}. They
included the Kamiokande (Kamioka Nucleon Decay Experiment) and IMB
(Irvine–Michigan–Brookhaven) detectors. They are large water
$\rm\check{C}$erenkov detectors of kiloton pure water with $O(10^3)$
PMTs. At that time, the GUT scale $(M_{\rm GUT})$ in the minimal SU(5)
GUT was expected to be around $10^{14}$~GeV from the unification of
the three gauge coupling constants in the SM, though the uncertainties
of the measured gauge coupling constants were large
\cite{langacker_rev}. The proton lifetime was predicted to be $\sim
10^{30}$~years. The experiments gave upperbounds on the partial
lifetime\footnote{The partial lifetime of the process is inverse of
  the event rate of the process, or the proton lifetime over the
  branching ratio.}  of $p\rightarrow \pi^0{\rm e}^+$ larger than
$\sim 10^{32}$~years, and the minimal SU(5) GUT was considered to be
excluded around 1986 \cite{theory86}.

The minimal SU(5) GUT is the simplest model in the GUTs, though it has
several problems. The gauge hierarchy problem is one of them
\cite{GH}. The mass term of the Higgs boson in the SM suffers from the
quadratic divergence in the radiative correction. The gauge hierarchy
between the weak and GUT scales is unstable even if we fine-tune
parameters in the minimal SU(5) GUT at tree level. One of the
solutions is introduction of supersymmetry (SUSY)
\cite{SUSYGUT}. Superpartners are introduced for each particle in
supersymmetric theories, and the radiative corrections to the scalar
mass terms are at most logarithmically divergent. The gauge hierarchy
is stable in the supersymmetric models. The supersymmetric extensions
of the GUTs and SM are the SUSY GUTs and the SUSY SM, respectively
\cite{SUSY}. The SUSY breaking scale ($m_{\rm SUSY}$) corresponds to
the masses of superpartners in the SUSY SM. The SUSY breaking scale
was expected to be below $O(1)$~TeV, since additional fine-tuning in
the SUSY SM, the {\it little hierarchy} between the SUSY breaking and
the weak scales, was not hopeful. It was pointed out that the GUT
scale raised to $\sim 10^{16}$~GeV in the SUSY SU(5) GUTs, since the
beta functions of the gauge coupling constants are changed due to the
introduction of superpartners in the SUSY SM.

In 1990s, the Weinberg angle was precisely measured in the LEP and SLC
experiments. It was found that the three gauge coupling constants are
successfully unified at $\sim 2\times 10^{16}$~GeV when they are
extrapolated to the higher energy in the SUSY SM \cite{gcu}. This was
the big step to consider the SUSY GUTs would be more realistic beyond
the conceptional ideas. Since the test of gauge coupling unification
was successful, the next test was proton decay. Since the GUT scale
raised to $\sim 2\times 10^{16}$~GeV, the prediction of the partial
proton lifetime of $p\rightarrow \pi^0{\rm e}^+$, induced by the $X$
boson, is $10^{(35-36)}$~years \cite{HMY}. It was safe from the bounds
derived by the Kamiokande and IMB experiments. On the other hand, the
SUSY GUTs have another source of proton decay.

The supersymmetric extensions of SM and GUTs introduce scalar fields
with baryon or lepton numbers, squarks and sleptons, as superpartners
of quarks and leptons, respectively. Thus, we can introduce
baryon-number violating dimension-five operators \cite{dimen5,
  dimen51}. In the minimal SUSY SU(5) GUT, the operators are generated
by colored Higgs multiplets, which are SU(5) partners of the Higgs
bosons and Higgsino in the SUSY SM. The operators are proportional to
inverse of the colored Higgs mass ($M_{H_C}$). The squarks and
sleptons in the operators are heavier than proton mass, and then the
dimension-six operators are generated by exchange of gauginos or
Higgsino in the SUSY SM.
Though only the gaugino exchange diagrams
were evaluated in the earlier papers \cite{dimen5gaugino,HMY}, it was
pointed out by Ref.~\cite{dimen5higgsino} that the Higgsino exchange
diagram may be comparable to the gaugino ones. The proton lifetime is
proportional to $M_{H_C}^2 m_{\rm SUSY}^2$. The dimension-five
operators are more dangerous than the dimension-six ones. The proton
decay rates are suppressed by the Yukawa coupling constants in the
minimal SUSY SU(5) GUT. However, the predicted partial lifetime of
$p\rightarrow K^+{\bar{\nu}}$ is smaller than $10^{32}$~years in the
minimal SUSY SU(5) GUT, assuming $m_{\rm SUSY}<O(1)$~TeV.  The decay
mode $p\rightarrow K^+{\bar{\nu}}$ is the largest in the minimal SUSY
SU(5) GUT unless it is accidentally suppressed.

The SuperKamiokande experiment started on 1996. The detector is a large water $\rm\check{C}$erenkov detector of 50 kilotons pure water with $O(10^4)$ PMTs. The sensitivities to proton decay were quite enhanced. The experiment updated upperbounds on partial lifetimes of various proton decay modes. The current bounds on the partial lifetimes $p\rightarrow \pi^0{\rm e}^+$ and $p\rightarrow K^+\bar{\nu}$ are $2.4\times 10^{34}$~years \cite{SKpie} and $6.6\times 10^{33}$~years \cite{SKknu}, respectively. When the Higgsino exchange diagrams were included in the analysis of the dimension-five operator proton decay in 1999, it was found that the accidental suppression of the proton decay did not happen and that that the minimal SUSY SU(5) GUT was also excluded \cite{dimen5higgsino}.

The minimal SUSY SU(5) GUT was excluded, though  the success of the gauge coupling unification in the SUSY GUTs was still quite beautiful. People believed that some mechanism would suppress the dimension-five operator proton decay. The introduction of some symmetries, such as the Peccei-Quinn symmetry \cite{dimen5,HMTY} or the $R$ symmetry in the extra-dimensional models \cite{HN}, was proposed to suppress it.

In 2008, the LHC experiments started. The superpartners in the SUSY SM and the Higgs boson in the SM were expected to be discovered at the experiments before the experiments started. People guessed that the squarks might be discovered earlier than the Higgs boson. However, the fact was the opposite. The Higgs boson was discovered in 2012 \cite{higgs}, while the superpartners have not yet been discovered \cite{SUSYsearch}. In addition, the observed Higgs boson mass is around 126~GeV. It is too heavy to be predicted in the SUSY SM if the stops, the superpartners of top quark, are lighter than a few TeV \cite{higgsmassinmssm}. It is consistent with the null result of superpartner searches at LHC.

The superpartner masses might be much heavier than $O(1)$~TeV even if they exist. However, since people had considered that the SUSY breaking scale is below $O(1)$~TeV from a viewpoint of the little hierarchy problem, this fact gave a big impact on studies of physics beyond the SM (BSM). One choice is to abandon SUSY. Another choice is to accept a possibility that the squarks and sleptons might have masses heavier than the weak scale \cite{splitsusy}. Some people consider the non-SUSY GUTs again. But, in this review, I follow the second choice, called the mini-split SUSY model, assuming the squark and slepton masses are around $O(10^{(2-3)})$~TeV \cite{minisplit, minisplit1}. While we have to abandon SUSY as a solution to the little hierarchy problem, we reserve the other advantages of the SUSY GUTs, such as the gauge coupling unification, the WIMP dark matter, and the gauge hierarchy problem. In addition, the various issues in the SUSY SM are solved in the model. They include the dimension-five operator proton decay. The dimension-five operator proton decay is suppressed by raising squark and slepton masses, and even the minimal SUSY SU(5) GUT is revived.

Now new experiments for proton decay are being prepared for. The main target of the HyperKamiokande, DUNE, and JUNO experiments is studies of the neutrino oscillation, though they have higher sensitivities to the proton decay searches. The proton decay may be discovered there.

In 2002, a meeting of theorists and experimentalists in particle physics was held at International Institute of Advanced Studies, Japan with title ``Origin of Matter'' \cite{nagashima_conf}. Prof. Koshiba gave a summary talk of the meeting. He said ``I have to give candid advice to you, especially theorists. Why are you still studying GUTs and SUSY, which have been already excluded ?'' I guess that he might consider that GUTs and SUSY were excluded by the Kamiokande and SuperKamiokande experiments.  At the time and now, theorists still have reasons to consider SUSY and GUTs and hope that proton decay will be searched for furthermore.

In the next section I give a general review of baryon-number violating nucleon decay, the current experimental bounds on proton decay, and the future prospects. In Section~\ref{sec3}, I discuss dimension-six and five operator proton decay in the mini-split SUSY model after introducing it. Section~\ref{sec4} is devoted to a summary of this review. In this review, we assume SUSY GUTs in 4-dimensional spacetime, but not non-SUSY GUTs and GUTs in extra-dimensional space since my space and time are limited.

\section{Baryon-number violating nucleon decay processes and the future prospects}

Rare processes are sensitive to BSM, which the energy frontier experiments are difficult to access. Among them, the baryon-number violating nucleon decays are the most sensitive to physics beyond the SM at an extremely high energy scale. Assuming that there is no lighter fermion than proton except for leptons, the processes are classified as follows.
\begin{description}
\item[$\Delta(B+L)=2$ nucleon decay: $p\rightarrow \pi^0{\rm e}^+$, $n\rightarrow \pi^0\bar{\nu} $, $\cdots$] \mbox{}\\
  The lowest effective operators are dimension-six in the SM. The list of the operators is given in Ref.~\cite{dim6list}. Some of them are predicted in various GUT models. The various modes are being updated by the SuperKamiokande experiment \cite{pdg}.  The future prospects are presented below. 
\item[$\Delta(B-L)=2$ nucleon decay: $n\rightarrow \pi^+e^-$, $\cdots$]\mbox{}\\ 
  They are induced by the dimension-seven effective operators in the SM, and they include the SM Higgs boson or derivative. The list of the operators is given in Ref.~\cite{dim7list}.  They are predicted in some SO(10) GUTs with the intermediate scale. They might be linked to baryogenesis, since the $B$-$L$ number generated in the early universe is not be washed out by the sphaleron process \cite{BM}. The charge discrimination is required to identify the  $\Delta(B-L)=2$ modes. Then, the bounds on them have not been updated since 1990s.
   
\item[$\Delta B=2$ dinucleon decay: $pp\rightarrow \pi^+\pi^+$, $\cdots$]\mbox{}\\
  The processes are possible inside nucleus. They are induced by the dimension-nine effective operators in the SM. It is pointed out that some SO(10) GUTs with the intermediate scale predict them, and that they may be linked to Majorana neutrino masses.
  The lower limit of lifetime for neutron bound in $^{16}$O from the SuperKamiokande experiment
  gives the bound on the $n$-$\bar{n}$ oscillation as $\tau(n$-$\bar{n})>4.7 \times 10^8~{\rm sec}$ \cite{dinucleon}.  It is better than the latest bound from the free neutron experiment $\tau(n$-$\bar{n})>8.6 \times 10^7~{\rm sec}$ \cite{nnbar}.
\end{description}  

The SuperKamiokande experiment gives upperbounds on various decay modes mentioned above. The $\Delta(B-L)=2$ or $\Delta B=2$ processes might be interesting since they are predicted in some SO(10) GUTs with the intermediate scale. But, the successful gauge coupling unification suggests SUSY GUTs without the intermediate scale. Thus, we concentrate on the $\Delta(B+L)=2$ nucleon decay, $p\rightarrow \pi^0{\rm e}^+$ and $p\rightarrow K^+\bar{\nu}$. They are induced by the dimension-six and five operators in the SUSY SU(5) GUTs. The baryon-number violating neutron decay processes are also being searched for, though the bounds are comparable to or weaker than  the proton decay processes of the counterparts \cite{pdg}.

The current bounds on the processes are derived by the SuperKamiokande
experiments as $\tau(p\rightarrow \pi^0{\rm e}^+) > 2.4\times 10^{34}$~years \cite{SKpie} and $\tau(p\rightarrow K^+\bar{\nu}) > 6.6 \times 10^{33}$~years \cite{SKknu}. The $\pi^0{\rm e}^+$ mode can be fully constructed by
the water $\rm\check{C}$erenkov detector, and the backgrounds are
suppressed. The experimental sensitivities are determined by the fiducial
volume of the detector. On the other
hand, the $K^+\bar{\nu}$ mode has a missing momentum, and it suffers
from the atmospheric neutrino background. It is necessary to
understand the atmospheric neutrinos in order to study the
$K^+\bar{\nu}$ mode. 

The HyperKamiokande detector \cite{HK} is now being constructed. It is also a
water $\rm\check{C}$erenkov detector of 187 kilotons pure water. Due to
the larger volume, it has the best prospect on $p\rightarrow
\pi^0{\rm e}^+$. However, due to the larger statistics, more dedicated
studies are needed to suppress the atmospheric neutrino background. It
is found by the simulation that the 90$\%$ CL limit may reach 
$7.9~(13)\times 10^{34}$~years for 10 (20)~year run, assuming a second tank is
installed after 6~years.

For the $K^+\bar{\nu}$ mode, the JUNO \cite{JUNO} and DUNE experiments \cite{DUNE} would be
competitive to the HyperKamiokande experiments \cite{HK} even though their
fiducial values are smaller. They are the liquid scintillator
detectors so that they have better energy resolution to reject the
atmospheric neutrino backgrounds. The prospects of the 90$\%$ CL limit
of the JUNO, DUNE and HyperKamiokande experiments are $1.9(4.0)\times
10^{34}$~years, $3.3(6.5)\times 10^{33}$~years, and $3.2(5.0)\times
10^{34}$~years for 10 (20) years run, respectively.

\section{Proton decay in SUSY SU(5) GUTs}
\label{sec3}

Now we discuss proton decay in SUSY GUTs. Before going to it, we
review the mini-split SUSY model, which is consistent with the current
experimental data, and discuss the GUT-scale mass spectrum expected from the
gauge coupling unification in the model.

\subsection{Mini-split SUSY model}

The SUSY SM is motivated by {\it 1)} the gauge hierarchy problem between the weak and GUT scales, {\it 2)} the successful gauge coupling unification, compatible with SUSY GUTs, and  {\it 3)}
the stable lightest SUSY particle (LPS) due to the R parity, which is  a candidate of the  dark matter in the universe. However, It has been known that the SUSY SM with $m_{\rm SUSY}<O(1)$TeV has some problems since the supersymmetric extension of the SM was introduced.  They are listed as follows.
\begin{description}
\item[(1) The FCNC and CP problems:]\mbox{}\\ While the flavor-violating and/or CP-violating SUSY breaking terms are allowed in the Lagrangian, they are constrained by FCNC and CP-violating processes, such as $K^0$-$\bar{K}^0$ mixing and electric dipole moments \cite{flavor}. If the off-diagonal terms are not suppressed in the squark mass matrices compared with the diagonal ones, the squark masses are larger than $O(10^{(2-3)})$~TeV.
\item[(2) The gravitino problem:] \mbox{}\\Gravitino is the superpartner of gravitino with spin $3/2$. It has only gravitational interactions suppressed by the Planck scale. If it is unstable, the lifetime is given by $\tau_{3/2}
  \sim
  10~{\rm sec}
  \times
  (10{\rm TeV}/m_{3/2})^3$
  ($m_{3/2}$ is the gravitino mass). If the gravitino mass is $O(1)$~TeV, the successful nucleosynthsis may be spoiled \cite{gravitino}.
\item[(3) The dimension-five operator proton decay problem:] \mbox{}\\The baryon-number violating dimension-five operators are induced in the SUSY GUTs as mentioned above. In addition, the operators suppressed by the Planck scale may be generically present, though they are constrained by the proton decay searches. They may be suppressed by some global symmetries while global symmetries are argued to be incompatible with quantum gravity.
\end{description}

It was considered that these problems would be solved by some SUSY breaking mechanisms or some cosmological scenarios. For example, people constructed many models of SUSY breaking to generate the universal SUSY breaking scalar mass terms with state-of-art technology of quantum field theories. Now a new problem has appeared. That is the observed Higgs boson mass. In the minimal SUSY SM, the Higgs boson mass is smaller than the $Z$ boson mass at tree level, and the radiative correction from the top quark and stop loops raises it, but not to 126~GeV if the stops are lighter than O(1) TeV \cite{higgsmassinmssm}.

Now we introduce the mini-split SUSY model \cite{minisplit,minisplit1}. In this model, the SUSY breaking terms are generated by the gravity mediation, and the gravitino mass is the order parameter in the SUSY breaking in the supergravity \cite{SUSY}. The gravitino mass is $O(10^{(2-3)})$~TeV, and  the SUSY breaking scalar masses are also comparable to the gravitino mass. Then, the problems mentioned above are automatically solved in this setup. The gaugino masses are generated by the anomaly mediation mechanism \cite{anomalymed} so that they are suppressed by one-loop factors compared with the gravitino mass. The gaugino masses are proportional to the beta functions of gauge coupling constants. Then, the SU(2)$_L$ gaugino, called as wino, is the lightest among the gauginos. The Higgsino mass would be comparable to the gravitino mass. But, it could be lighter if the Higgsino mass is suppressed by some symmetry.

Wino or Higgsino is the LSP in the mini-split SUSY model, and they are the dark matter candidates. If they are coupled with the thermal bath in the early universe and they explain the observed dark matter abundance, the wino mass is about 3~TeV \cite{winomass} and the Higgsino mass is 1~TeV \cite{higgsinomass}. Since they have the electroweak interactions, their masses are $O(1)$~TeV, not $O(100)$~GeV. 

The difference among the beta functions of tree gauge coupling constants in the SUSY SM comes from the gauginos and  the Higgsino in addition to the gauge and Higgs bosons. Since the gauginos are $O(1)$~TeV, the successful unification of the gauge coupling constants is not spoiled. Furthermore, if the Higgsino mass is $O(10^{(2-3)})$~TeV, the gauge coupling unification is improved, as discussed below \cite{ggc_split}.

The little hierarchy between the weak and SUSY breaking scales would be accidental in the mini-split SUSY model. The parameters in the Higgs potential must be finely tuned so that the Higgs vacuum expectation value is much smaller than the SUSY breaking scale. However, the big hierarchy between the  SUSY breaking  and GUT scales is still stabilized by supersymmetry.

Last, we have an additional advantage of the mini-split SUSY model. The model building of the SUSY breaking becomes much easier due to the anomaly mediation, which is built in supergravity. What we need is only the introduction of the hidden sector decoupled from the SUSY SM, which is responsible for the spontaneous SUSY breaking \cite{minisplit, minisplit1}. The supergravity generates the hierarchical SUSY breaking masses for scalar bosons and gauginos. It is the reason why ``mini-split SUSY'' is called ``pure-gravity mediation'' in Ref.~\cite{minisplit}.

The possible experimental tests of the mini-split SUSY model are limited since squarks and sleptons are quite heavy. The LSP, wino or Higgsino, is a window to the model. The future prospects of the direct \cite{directwino} and indirect dark matter searches \cite{hmn}  may reach the prediction of the wino and Higgsino dark matter. 

Another window is violation of fundamental symmetries. The electric dipole moment of electron is induced by the Barr-Zee two-loop diagrams in the Higgsino-wino system, and the prediction may be within the future prospects of the experiments \cite{nakai}. Proton decay might be another one. The minimal  SUSY SU(5) GUT is revived in the mini-split SUSY model. The prediction may be reached by future experiments.

\subsection{GUT-particle mass spectrum from the gauge coupling unification}

The gauge coupling unification gives constraints on the GUT-particle mass spectrum \cite{hmyg}. First, let us assume the minimal SUSY SU(5) GUT for simplicity\footnote{The particle contents and Lagrangian of the minimal SUSY SU(5) GUT are given in Ref.~\cite{HMY}.}.
In this case, the GUT-particle mass spectrum includes
the $X$ boson mass ($M_X$), the GUT breaking Higgs mass ($M_\Sigma$), and the colored Higgs mass ($M_{H_C}$). The GUT gauge symmetry is broken by the vacuum expectation value of the SU(5) adjoint Higgs multiplet. The three gauge coupling constants in the SM are related to the SU(5) one via the renormalization-group equations. Then, we get two relations among three gauge coupling constants after removing the SU(5) gauge coupling constant,
\begin{eqnarray}
  \frac{3}{\alpha_2(m_Z)}
  -\frac{2}{\alpha_3(m_Z)}
  -\frac{1}{\alpha_1(m_Z)}
  &=&
  \frac{1}{2\pi}
  \left[
    \frac{12}{5}\ln\left(\frac{M_{H_c}}{m_Z}\right)
    -\frac{2}5\ln\left(\frac{M^4_{\tilde{H}}M_{H}}{m^5_Z}\right)
   +4\ln\left(\frac{m_3}{m_2}\right)
   \right],
  \nonumber\\
  \frac{5}{\alpha_1(m_Z)}
  -\frac{3}{\alpha_2(m_Z)}
    -\frac{2}{\alpha_3(m_Z)}
  &=&
  \frac{1}{2\pi}
  \left[
    {12}\ln\left(\frac{M_{X}^2M_{\Sigma}}{m_Z^3}\right)
   -2\ln\left(\frac{m_2}{m_Z}\right)
   +4\ln\left(\frac{m_3}{m_Z}\right)
   \right].
\label{RGE}
\end{eqnarray}
Here, $\alpha_i(m_Z)$ ($i=$1-3) are the gauge coupling constants in the SM at the $Z$ boson mass ($m_Z$). The U(1) gauge coupling constant is given by $\alpha_1=5\alpha_Y/3$. The factor $5/3$ in $\alpha_1$ comes from the GUT normalization.  $m_{\tilde{H}}$, $m_H$, $M_3$, and $M_2$ are masses of Higgsino, the second Higgs boson, gluino and wino, respectively,  in the SUSY SM. We take a common mass for squarks and sleptons for simplicity so that their dependences disappear in the above equations. The first equation comes mainly from the splitting between masses of the Higgs bosons and Higgsino in the SUSY SM and the colored Higgs mass, which are embedded into the SU(5) ${\bf 5}$ and  ${\bar {\bf 5}}$ multiplets, The second one is from the mass splitting in the gauge and GUT breaking Higgs multiplets. It is found that the colored Higgs mass is constrained from the gauge coupling constant unification. On the other hand, the $X$ boson mass is not determined since it is correlated with the GUT breaking Higgs mass. 

 The gaugino masses in the SUSY SM are generated by the anomaly mediation as \cite{anomalymed}
\begin{eqnarray}      
M_1&=&\frac{33}3\frac{\alpha_1}{4\pi}\left(m_{3/2}+\frac{1}{11}L\right),\nonumber\\
M_2&=&\frac{\alpha_2}{4\pi}\left(m_{3/2}+L\right),\nonumber\\
M_3&=&-3\frac{\alpha_3}{4\pi}m_{3/2},
\end{eqnarray}
where $L$ comes from the Higgsino and the Higgs boson one-loop diagrams,
\begin{eqnarray}
  L&=&m_{\tilde{H}}\sin 2\beta \frac{m_H^2}{|m_{\tilde{H}}|^2-m_{{H}}^2}\log\frac{|m_{\tilde{H}}|^2}{m_{{H}}^2}, \nonumber
\end{eqnarray}
with $\tan\beta$ the vacuum expectation value ratio of two Higgs bosons in the SUSY SM. $L$ is not negligible if the Higgsino mass is $O(10^{(2-3)})$~TeV. In the mini-split SUSY model, $m_{H}$, is $O(10^{(2-3}))$~TeV. We take the ratios of the gaugino masses as a free parameters.

\begin{figure}[!h]
\centering\includegraphics[width=3.0in]{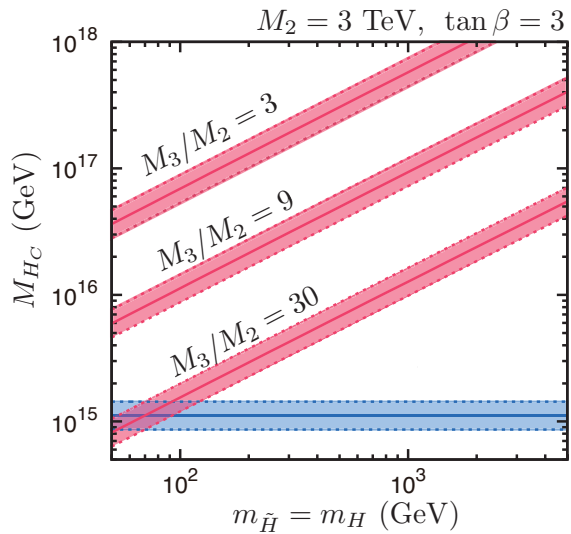}
\caption{Colored Higgs mass, $M_{H_C}$, evaluated from gauge coupling unification, as a function of $M_{\tilde{H}}=M_H$ (red regions). Here, wino mass is 3~TeV and $\tan\beta=3$. Gluino and wino mass ratio is taken to be 3, 9, 30.
  Blue region is prediction of SUSY SM with $m_{\rm SUSY}<O(1)$~TeV. Original of this figure is given in Ref.~\cite{ggc_split}.
}
\label{mhcfig}
\end{figure}

In Fig.~\ref{mhcfig} we show the colored Higgs mass $M_{H_C}$, evaluated from the gauge coupling unification, as a function of $M_{\tilde{H}}=M_H$ (red regions).
The original of this figure is given in Ref.~\cite{ggc_split}. Here, the wino mass is 3~TeV, and $\tan\beta=3$. The width of the region comes mainly from uncertainty of $\alpha_3(m_Z)$. The gluino and wino mass ratio is taken to be 3, 9, 30. The blue region is the prediction of the SUSY SM with $m_{\rm SUSY}<O(1)$~TeV. It is pointed in Ref.~\cite{hmyg} that the colored Higgs mass is lower than the GUT scale ($\sim 10^{16}$~GeV) in the SUSY SM with $m_{\rm SUSY}<O(1)$~TeV. In the mini-split SUSY model, the colored Higgs mass could be predicted around the GUT scale. This implies that the gauge coupling unification could be more precisely realized in the model. 

\begin{figure}[!h]
\centering\includegraphics[width=3.0in]{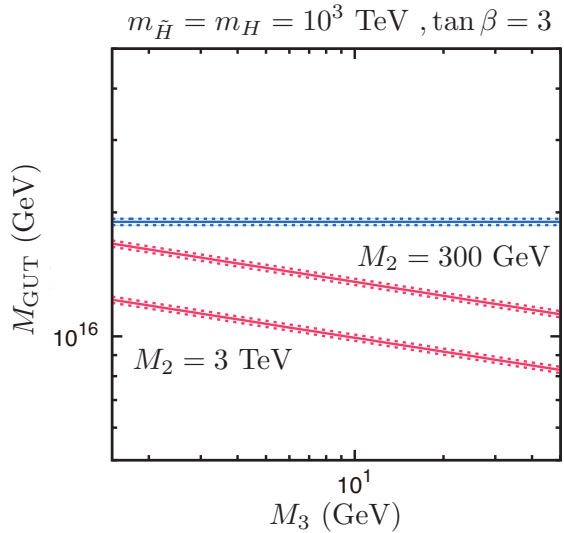}
\caption{$M_{\rm GUT}\equiv (M_X^2M_\Sigma)^{1/3}$, evaluated from gauge coupling unification, as a function of gluino mass, $M_3$. Here, we take $M_{\tilde{H}}=M_H=10^3$~TeV, $\tan\beta=3$, and $M_2=300$~GeV and $3$~TeV. Blue region  is prediction of SUSY SM $m_{\rm SUSY}<O(1)$~TeV. Original of this figure comes from Ref.~\cite{ggc_split}.
}
\label{mgutfig}
\end{figure}

Next, we define the GUT scale as $M_{\rm GUT}\equiv (M_X^2M_\Sigma)^{1/3}$, and  we show $M_{\rm GUT}$ as a function of the gluino mass,
$M_3$ in Fig.~\ref{mgutfig}. The original of this figure is also given in
Ref.~\cite{ggc_split}. Here, we take $M_{\tilde{H}}=M_H=10^3$~TeV,
$\tan\beta=3$, and $M_2=300$~GeV and $3$~TeV. The blue region is the
prediction of the SUSY SM with $m_{\rm SUSY}<O(1)$~TeV. The uncertainty on
the prediction from $\alpha_3(m_Z)$ is much smaller than that of the
colored Higgs mass.

When the gaugino masses are heaver, $M_{\rm GUT}$ is lower. We cannot determine the $X$ boson mass from the gauge coupling unification, since the $X$ boson mass is correlated with the GUT breaking Higgs mass in $M_{\rm GUT}$. However, the lower GUT scale might suggest the enhancement of the dimension-six proton decay.

Now we discuss the model dependence of the prediction of the GUT-particle mass spectrum. In the minimal SUSY SU(5) GUT, the SU(5) gauge symmetry is broken by the SU(5) adjoint Higgs multiplet. The large splitting between the colored Higgs mass and masses of the Higgsino and Higgs bosons in the SUSY SM is given by hand. This tree level fine-tuning is not avoidable in the model, though the fine-tuning is not spoiled by the radiative correction due to supersymmetry. This problem is called as the doublet-triplet splitting problem. The missing partner model \cite{missingmodel} is introduced as the solution to the problem\footnote{The particle contents and lagrangians of the missing partner model and the Peccei-Quinn symmetric extension are given in Ref.~\cite{HMTY}.}. In the model, the SU(5) gauge symmetry is broken by the {\bf 75} dimensional Higgs multiplet, and ${\bf 50}$ and ${\bf \bar{50}}$ are also introduced for the mechanism to work. In this model, the mass splitting among the components in the {\bf 75} dimensional Higgs multiplet contributes to the difference of the three gauge coupling constants in the SM. As the result, the {\it effective} colored Higgs mass, which appears in the dimension-five operators, is $2\times 10^4$ times larger than the prediction of the minimal SUSY SU(5) GUT \cite{HY}. The GUT scale $M_{\rm GUT}$ is lower than that of the minimal model by only 1.4.

The {\bf 75}, and  ${\bf 50}$ and ${\bf \bar{50}}$ dimensional Higgs multiplets give big contributions to the beta function of the SU(5) gauge coupling constant. Especially, the SU(5) gauge coupling constant blows up above the mass of ${\bf 50}$ and ${\bf \bar{50}}$. It is suggested from the gauge coupling unification that the effective colored Higgs mass is much larger than the GUT scale, though the pertaurbative picture is broken in the minimal model. If the Peccei-Quinn symmetry is introduced and  pairs of ${\bf 50}$ and  ${\bf \bar{50}}$ and
${\bf 5}$ and  ${\bf \bar{5}}$ Higgs multiplets are introduced, the pertaurbative picture is not broken and the dimension-five operator proton decay is suppressed \cite{HMTY}.

The minimal SUSY SU(5) GUT and the missing partner model are simple models. However, unknown particles with GUT scale masses may be introduced and the mass splitting in the GUT multiplets may be generated due to their coupling to the GUT breaking Higgs field. In this case, they may contribute to the differences among the gauge coupling unification, and the above results of the GUT-particle mass spectrum may be a merely qualitative expectations from the gauge coupling unification.

\subsection{Dimension-six operator proton decay}

\begin{figure}[!h]
\centering\includegraphics[width=2.0in]{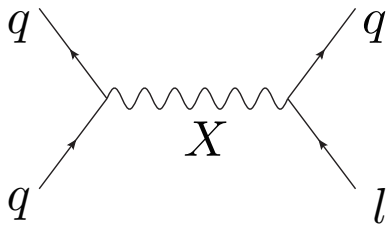}
\caption{Dimension-six operators induced by X boson. }
\label{xboson}
\end{figure}

The baryon-number violating dimension-six operators are induced by the $X$ boson in the SUSY SU(5) GUTs (Fig.~\ref{xboson}). The proton lifetime is given as
\begin{eqnarray}
  \tau(p\rightarrow \pi^0{\rm e}^+)\simeq 1 \times 10^{35}\times \left(\frac{M_X}{10^{16}{\rm GeV}}\right)^4 {\rm years},
\end{eqnarray}
where $M_X$ is the $X$ boson mass. Thus, if $M_X<10^{16}$~GeV, the HyperKamiokande may discover the $X$ boson proton decay. The GUT scale might be smaller than $10^{16}$~GeV in the mini-split SUSY model as mentioned above, though the gauge coupling constant unification cannot determine the $X$ boson mass itself.

The prediction for the proton lifetime of the $X$ boson is precisely evaluated now if explicit models of the SUSY GUT are assumed. The procedure for the evaluation is following. {\it 1)}  The effective dimension-six operators are derived by integrating the $X$ boson. The matching conditions are derived at one-loop level \cite{PDmatching}. {\it 2)} The Wilson coefficients of the operators at the hadronic scale are evaluated with the RGEs at two-loop level \cite{HKMN}\cite{NA}. The threshold corrections to them at the SUSY breaking scale may be included though it is negligible \cite{PDmatching}. {\it 3)} Using the hadronic matrix element evaluated by the lattice QCD, the proton decay rate is evaluated. The latest lattice QCD results for the matrix elements of proton decay is given by Ref.~\cite{yaoki}, in which the matrix elements are evaluated at physical pion mass. The proton decay of the $X$ boson is less sensitive to the detail of the GUT model if the $X$ boson mass is assumed. However, the model dependence appears through the matching condition of the operators at one loop level. The correction to the proton decay rate from the one-loop matching in the minimal SUSY SU(5) GUT is about 5\% \cite{PDmatching}.  In the missing partner model with Peccei-Quinn symmetry, the correction is $O(10)\%$ \cite{bhko}. This is because the model has larger SU(5) multiplets, such as  ${\bf 75}$.

Another model dependence comes from the low-energy model below the GUT scale. We assumed that the low-energy model is the SUSY SM with three generations. However, the model might have additional matters. For example, ${\rm E}_6$ GUTs or string theories may introduce extra matter families. If they form SU(5) multiplets, the gauge coupling unification is not spoiled. The introduction of the additional matters increases the SU(5) gauge coupling constant, and the proton decay rate is enhanced \cite{HKN_extra}. 

\begin{figure}[!h]
\centering\includegraphics[width=4.0in]{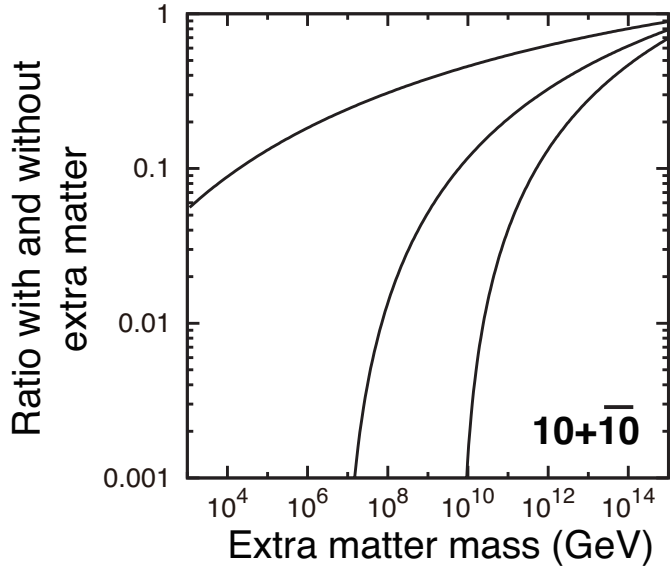}
\caption{
Ratio of proton lifetime with and without  extra matter as functions of extra matter  mass of (${\bf 10}+{\bf \bar{10}}$). Here, proton decay is induced by  $X$ boson. Here, 1,2,and 3 pairs of ${\bf 10}+{\bf \bar{10}}$ are introduced from top to bottom lines, respectively. Original of this figure comes from Ref.~\cite{HKN_extra}.
}
\label{enhancement}
\end{figure}

In Fig.~\ref{enhancement}, we show the ratio of the proton lifetime with and without the extra matter as a function of the extra matter mass. Here, we introduce SU(5) ${\bf 10}+{\bf \bar{10}}$ multiplets. We introduce 1,2,and 3 pairs of ${\bf 10}+{\bf \bar{10}}$ from the top to bottom lines, respectively in the figure. It is found that the proton lifetime becomes much shorter when the extra matters are introduced. 

In this review the SUSY SU(5) GUT is assumed. It is motivated by the successful gauge coupling unification. If proton decay is discovered and the branching ratios of proton decay are measured, we could get more information of the model behind the discovery. In Ref.~\cite{MM}  the branching ratios in the dimension-six proton decays in SUSY SU(5), SO(10), and E$_6$ GUT models are evaluated. In Ref.~\cite{EGNNO} they compare perdition of flipped and unflipped SU(5) GUTs.

\subsection{Dimension-five operator proton decay}

Integrating out the colored Higgs multiplets, we get the baryon-number violating dimension-five operators. The effective superpotential for the operators is given as \cite{HMY}
\begin{eqnarray}
      W&=&
      \frac{1}{2M_{H_C}} f_{u_i}f_{d_l}V_{kl}^\star {\rm e}^{i \varphi_i}
        Q_i Q_j Q_k L_l
        +
        \frac{1}{M_{H_C}} f_{u_i}f_{d_l}V_{kl}^\star {\rm e}^{i \varphi_i}
          \overline{U}_i \overline{E}_j \overline{U}_k \overline{D}_l
\label{dim5ope}
\end{eqnarray}
where $i,j,k, l$ are for generations. $Q$ and $L$ are the chiral superfields of the SU(2)$_L$ doublet quark and lepton, respectively, and $\overline{U}$, $\overline{D}$, $\overline{E}$ are those of the SU(2)$_L$ singlet quarks and lepton, respectively. The phase factor ${\rm e}^{i \varphi_i}$ ($\varphi_1+\varphi_2+\varphi_3=0$) is extra phase factors generic in the SUSY SU(5) GUT. The Yukawa coupling constants $f_{u_i}$ and $f_{d_i}$ are related to the quark masses $m_{u_i}$ and $m_{d_i}$ as $m_{u_i}=f_{u_i}v \sin\beta$  and $m_{d_i}=f_{d_i}v \cos\beta$ ($v$ is the vacuum expectation value of the SM Higgs field.). The Yukawa coupling constants and the Kobayashi-Maskawa matrix $V$ in the above effective superpotential are given at the GUT scale.
\begin{figure}[!h]
\centering\includegraphics[width=2.5in]{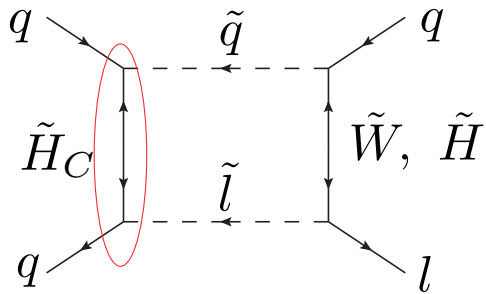}
\caption{
Proton decay induced by colored Higgs exchange. Red circle corresponds to dimension-five operator induced by colored Higgs exchange. Extra wino ($\tilde{W}$) or Higgsino ($\tilde{H}$) exchange is required for proton decay since squark ($\tilde{q}$) and slepton ($\tilde{l}$) are heavier than proton. 
}
\label{coloredhiggsfig}
\end{figure}

Due to the SU(3)$_C$ antisymmetric nature in Eq.~\ref{dim5ope}, we get $i\ne k$ in Eq.~\ref{dim5ope}. With the hierarchical structure of the Yukawa coupling constants, it favors the final states of proton decay to include the strange quarks. Thus, the main mode is $p\rightarrow K^+\bar{\nu}$. 

The dimension-five operators in Eq.~\ref{dim5ope} include squarks and/or sleptons, which are heavier than proton mass. The extra superpartner exchange is required for proton to decay. The gluino exchange is suppressed when the squark masses degenerate. The wino and Higgsino exchanges are dominant for the first and second operators in Eq.~\ref{dim5ope}, respectively (Fig.~\ref{enhancement}).

The dimension-six operators are generated by the wino/Higgsino exchange, and the Wilson coefficients are suppressed by $1/M_{H_C}m_{\rm SUSY}$. Exactly speaking, $m_{\rm SUSY}$ is a function of the superpartner mass spectrum. When the squark and slepton masses ($m_{S}$) are comparable to $M_2$ and $m_{\tilde{H}}$, $m_{\rm SUSY}\simeq 2m_{S}$. On the other hand, when $m_{S}$ is much heavier than $M_2$ or $m_{\tilde{H}}$ as in the mini-split SUSY model, $m_{\rm SUSY}$ is given by $m_{S}^2/M_2$ or $m_{S}^2/m_{\tilde{H}}$. It comes from the chirality flip of wino or Higgsino. Thus, the dimension-five operator proton decay is more efficiently suppressed in the mini-split SUSY model. When $m_{\tilde{H}}\gg M_2$, the Higgsino exchange contribution is dominant in the amplitude. In such a case, the partial lifetime of $p\rightarrow K^+\bar{\nu}$ is approximately given as
\begin{eqnarray}
  \tau(p\rightarrow K^+\bar{\nu})
  &=&4\times 10^{35}\times \sin^42\beta
  \left(\frac{0.1}{\bar{A}_R}\right)^2
  \left(\frac{M_S}{10^2~{\rm TeV}}\right)^2
      \left(\frac{M_{H_C}}{10^{16}~{\rm GeV}}\right)^2
           {\rm years}.
           \end{eqnarray}
Here, we take $m_S=m_{\tilde{H}}$. The symbol $\bar{A}_R$ is for the radiative correction, and the definition is given in Ref.~\cite{HKKN}. The factor $\sin^4 2\beta$ comes from both the colored Higgs and Higgsino exchanges. 

\begin{figure}[!h]
\centering\includegraphics[width=3.5in]{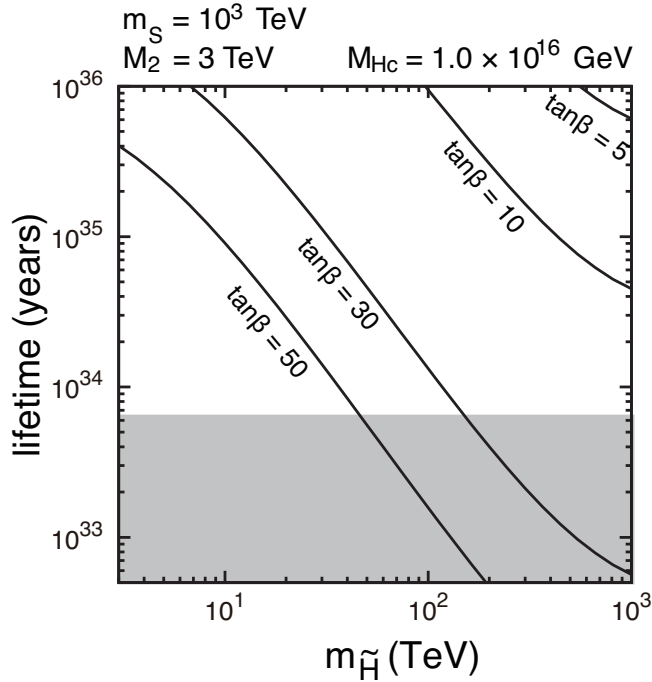}
\caption{
  Partial lifetime of $p\rightarrow K^+\bar{\nu}$ as a function of Higgsino mass in minimal SUSY SU(5) GUT. Wino mass is assumed to be 3~TeV, and squark and slepton masses are $10^3$~TeV, and  $M_{H_C}=1.0\times 10^{16}~{\rm GeV}$. Solid lines are for $\tan\beta=5,~,10~,30~$,and $50$ from right-top to left-bottom, respectively. Shaded region is excluded by current bound $\tau(p\rightarrow K^+\bar{\nu})>6.6\times 10^{33}$~years. Original of this figure is from Ref.~\cite{HKKN}.
}
\label{dim5fig2}
\end{figure}

In Fig.~\ref{dim5fig2} the partial lifetime of $p\rightarrow K^+\bar{\nu}$ is shown as
a function of the Higgsino mass. Here, $M_S=10^3$~TeV, $M_2=3$~TeV. $M_{H_C}=1.0\times 10^{16}$~GeV. Solid lines are for $\tan\beta=5,~,10~,30~$,and $50$ from right-top to left-bottom, respectively. The shaded region is excluded by the current bound, $\tau(p\rightarrow K^+\bar{\nu})>6.6\times 10^{33}$~years. As expected, the larger Higgsino enhances the proton decay rate. Notice that the lifetime is proportional to the fourth power of $m_S$ since the amplitude is proportional to $m_{\tilde{H}}/m_S^2$ for $m_{\tilde{H}}\ll m_S$.
It is found that the minimal  SUSY SU(5) GUT is still viable, and that the future experiments may find the dimension-five operator proton decay in the model.

In the above study of the dimension-five operator proton decay, we assume that the of-diagonal terms in the squark and slepton mass matrices are suppressed for simplify. They may be present in the mini-split SUSY model.  The rates and branching fractions of proton decay can be changed if they are introduced \cite{MS}.

\section{Summary of this review}
\label{sec4}

  I review proton decay in the SUSY SU(5) GUTs assuming the mini-split SUSY breaking model. In the mini-split SUSY breaking model, the squark and slepton masses are $O(10^{(2-3)})$~TeV while the gaugino masses are $O(1)$~TeV. As the result, even the minimal SUSY SU(5) GUT is still viable in the model. We present the motivation of the mini-split SUSY model and discuss the future prospects of proton decay searches in the SUSY SU(5) GUTs.

  Some serious issues to GUTs have appeared since it was proposed. But, new ideas appeared for the issues. The concepts of GUTs are still fascinating the researchers in particle physics. We hope that the next clues will appear to us. That is proton decay?

\section*{Acknowledgment}
This work was supported by JSPS Grant-in-Aid for Scientific Research
KAKENHI Grant Number JP21K03572. The work was also supported by World
Premier International Research Center Initiative (WPI Initiative),
MEXT, Japan.

\let\doi\relax

\end{document}